\documentclass[a4paper,11pt]{article}

\usepackage{jinstpub} 

\usepackage{subfigure}
\usepackage{siunitx}
\usepackage{footmisc}

\title{\boldmath Overview of CNM LGAD results: Boron Si-on-Si and epitaxial wafers}

\author[a,1]{C.~Grieco,\note{Corresponding author.}}
\author[a]{L.~Castillo Garc\'{i}a,}
\author[b]{A.~Doblas Moreno,}
\author[c]{E.~L.~Gkougkousis,}
\author[a,d]{S.~Grinstein,}
\author[b]{S.~Hidalgo,}
\author[b]{N.~Moffat,}
\author[b]{G.~Pellegrini,}
\author[b]{and J.~Villegas Dominguez}

\affiliation[a]{Institut de F\'{i}sica d'Altes Energies (IFAE), The Barcelona Institute of Science and Technology (BIST), \\Carrer Can Magrans s/n, Edifici Cn, Campus UAB, E-08193 Bellaterra (Barcelona), Spain}
\affiliation[b]{Centro Nacional de Microelectr\'{o}nica (CNM),\\Carrers dels Tillers, UAB Campus, E-08193 Bellaterra (Barcelona), Spain}
\affiliation[c]{Conseil Europ\'{e}en pour la Recherche Nucl\'{e}aire (CERN),\\ Esplanade des Particules 1, CH-1211 Geneva 23, Switzerland}
\affiliation[d]{Instituci\'o Catalana de Recerca i Estudis Avan{\c c}ats (ICREA), Passeig de Lluís Companys, 23, 08010 Barcelona, Spain}

\emailAdd{chiara.grieco@cern.ch}

\abstract{Low Gain Avalanche Detectors (LGADs) are n-on-p silicon sensors with an extra p-layer below the collection electrode which provides signal amplification. When the primary electrons reach the amplification region new electron-hole pairs are created that enhance the generated signal. The moderate gain of these sensors, together with the relatively thin active region, provide precise time information for minimum ionizing particles. To mitigate the effect of pile-up at the HL-LHC the ATLAS and CMS experiments have chosen the LGAD technology for the High Granularity Timing Detector (HGTD) and for the End-Cap Timing Layer (ETL), respectively. A full characterization of recent productions of LGAD sensors fabricated at CNM has been carried out before and after neutron irradiation up to 2.5 $\times$ 10$^{15}$ n$_{eq}$/cm$^{2}$ . Boron-doped sensors produced in epitaxial and Si-on-Si wafers have been studied. The results include their electrically characterization (IV and bias voltage stability) and performance studies (charge and time resolution) for single pad devices with a Sr-90 radioactive source set-up. The behaviour of the Inter-Pad region for irradiated 2 $\times$ 2 LGAD arrays, using the Transient Current Technique (TCT), is shown. The results indicate that the Si-on-Si devices with higher resistivity perform better than the epitaxial ones.}


\proceeding{The 12$^{\text{th}}$ International Conference on position Sensitive Detectors\\
  12$^{\text{th}}$-17$^{\text{th}}$ September 2021\\
  Birmingham, United Kingdom}

\begin{document}
\maketitle
\flushbottom

\section{Introduction}
\label{sec:intro}
In the past years, several doping materials (Boron\footnote{\label{note1}Production run 10478}, Boron plus Carbon\footref{note1} and Gallium doping\footnote{Production run 10924}) for LGAD sensors were explored by the CNM Barcelona~\cite{cnm_talk32rd50}. Their charge collection and time resolution as a function of the bias voltage at different temperatures were studied for both neutron and proton irradiation~\cite{past_prod}. Boron and Boron plus Carbon sensors perform similarly in terms of time resolution before and after irradiation (below \SI{50}{\pico\second} at higher voltages) but devices with Carbon collect more charge at the same bias voltage. Gallium presents 20\% less gain with respect to Boron doping and this research line has not been followed due to the poorer radiation hardness and timing performances. Carbon infusion seems to help more in diminishing the effect of gain reduction after irradiation. However, in this first run with Carbon the implantation process was not optimal and its benefits are not clear compared to later productions from other manufacturers.

\section{CNM LGAD productions}
More recently, CNM has produced two other LGAD runs with Boron doping for which the electrical characterization, the charge collection and timing performance are shown in this paper before and after neutron irradiation\footnote{Irradiation for all the sensors has been performed with \SI{1}{\mega\electronvolt} neutrons in the TRIGA reactor in Ljubljana.}. The first run on Si-on-Si wafers\footnote{Production run 12916, named AIDA2020.} with higher doping dose (1.8 $\times$ 10$^{13}$~atoms/cm$^{2}$) than the first Boron run mentioned in Section~\ref{sec:intro}. The second run is on low-resistivity epitaxial wafers\footnote{Production run 13002} with even a higher dose than the Si-on-Si devices (2 $\times$ 10$^{13}$~atoms/cm$^{2}$). The aim to investigate the usage of epitaxial wafers is their lower cost and the easier availability with respect to the Si-on-Si ones.

\section{Electrical characterization}
\begin{figure}[htbp]
\centering
\subfigure
	{\includegraphics[width=0.45\textwidth]{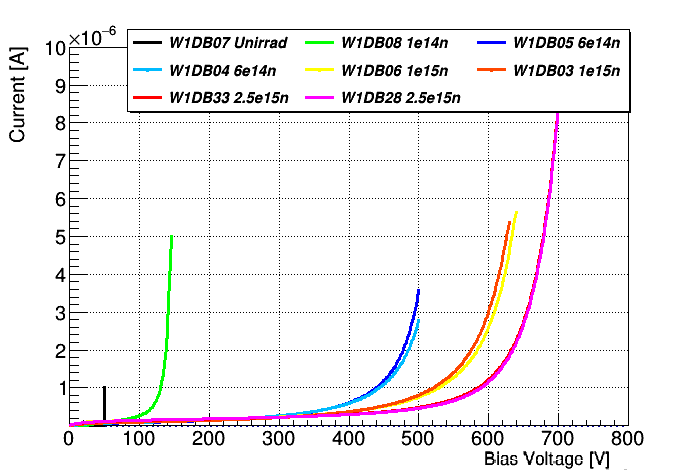}
	} \qquad
\subfigure
	{\includegraphics[width=0.45\textwidth]{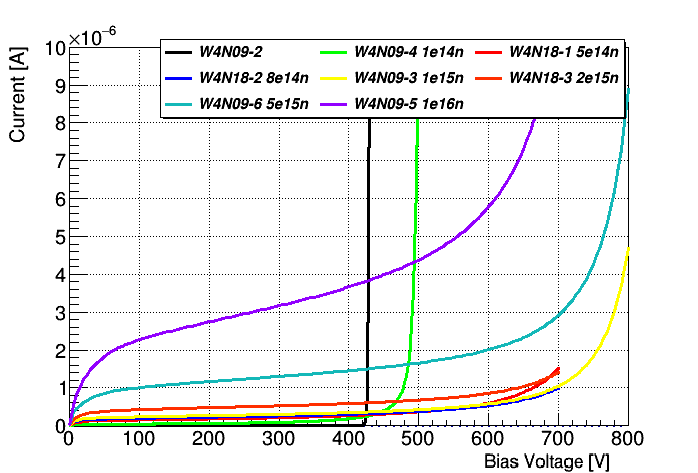}
} \qquad
\caption{IV curves on probe station for Si-on-Si (left) and low-resistivity epitaxial (right) wafer sensors before and after irradiation.}
\label{fig:IVs}
\end{figure}

Si-on-Si sensors irradiated up to 2.5 $\times$ 10$^{15}$ n$_{eq}$/cm$^{2}$ and low-resistivity epitaxial sensors irradiated up to 10$^{16}$ n$_{eq}$/cm$^{2}$ were electrically characterized in the CNM laboratory on a probe station at $-30~^{\circ}$C in a dry environment. The leakage current was studied as a function of the bias voltage, in order to determine the breakdown voltage, V$_{bd}$. In particular, gain layer and full depletion voltages, V$_{gl}$ and V$_{fd}$, are determined by C-V curves~\cite{albert_rd50}. The breakdown voltage measured after dicing is similar to the one measured at a wafer level~\cite{albert_rd50} despite the \SI{10}{\degree C} difference.
Sensors were biased with positive voltage on the pad and guard-ring, while the back of the sensor (ohmic side) was grounded. The plots in Figure~\ref{fig:IVs} show the total current I$_{tot}$=I$_{PAD}$+I$_{GR}$ for the sensors from Si-on-Si (left) and low-resistivity epitaxial (right) wafers before and after irradiation. In particular, for Si-on-Si sensors V$_{gl}\sim$\SI{38}{\volt}, V$_{fd}\sim$\SI{42}{\volt} and V$_{bd}\sim$\SI{50}{\volt}, and for epitaxial ones V$_{gl}\sim$\SI{30}{\volt}, V$_{fd}\sim$\SI{35}{\volt} and V$_{bd}\sim$\SI{400}{\volt} before irradiation. As expected, the bulk current and the breakdown voltage increase with the increasing of fluence for both productions. The breakdown voltage increases because of the lower quantity of dopant in the doping layer with respect to the less irradiated and unirradiated sensors~\cite{wiehe}.

\section{Stability measurements}
\begin{figure}[htbp]
\centering
\subfigure
	{\includegraphics[width=0.45\textwidth]{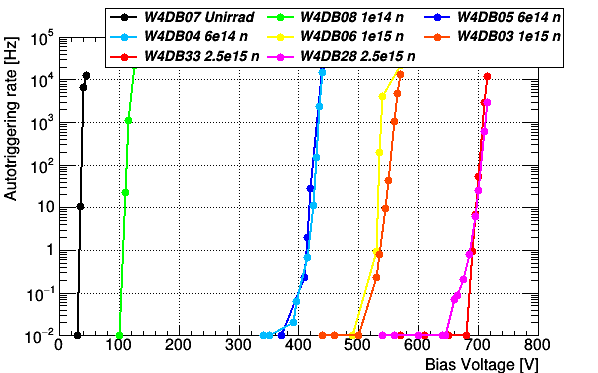}
	} \qquad
\subfigure
	{\includegraphics[width=0.45\textwidth]{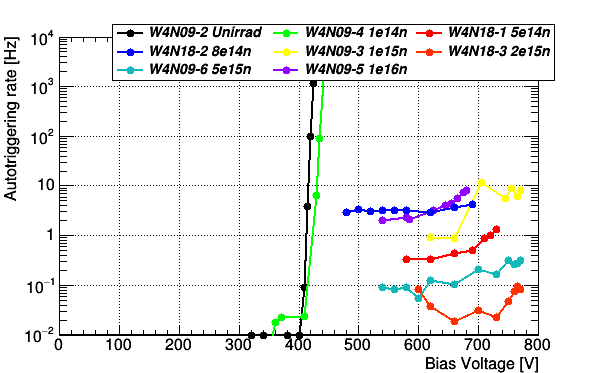}
} \qquad
\caption{Rate of self triggers measured at $-30~^{\circ}$C for Si-on-Si (left) and low-resistivity epitaxial (right) wafer sensors.}
\label{fig:autotriggering}
\end{figure}

An important parameter of the LGAD sensors is the voltage at which the rate of self induced triggers becomes large enough to prevent the operation of the devices, since this would generate a detector occupancy beyond the capability of the readout system. 
In the ATLAS High Granularity Timing Detector the rate of self triggers is limited to \SI{1}{\kilo\hertz}. 
The rate of self triggers for both Si-on-Si (left) and low-resistivity epitaxial (right) sensors is shown in Figure~\ref{fig:autotriggering}. The unirradiated sensor from the Si-on-Si wafer presents a high rate of self triggers which severely limits the operational range of the devices before irradiation, since, as stated in the previous section, the full depletion is of the order of \SI{42}{\volt}. However, after irradiation the operational range is extended.
On the other hand, for the devices from the epitaxial wafers the voltage for self triggers is much larger, and it is possible to operate the sensors before and after irradiation. Epitaxial sensors needs more voltage at the same fluence, for 10$^{14}$n$_{eq}$/cm$^2$ auto-triggers later than for Si-on-Si devices.

\section{Charge collection and time resolution}
\noindent
To study the performance of the sensors, coincidence triggers are acquired between the devices under test (DUTs) and a reference LGAD sensor with a Sr-90 radioactive source at \SI{-30}{\degree C} in a dry environment.
The set-up is the same shown in~\cite{past_prod} and the data are analyzed offline with the LGADUtils framework~\cite{LGADUtils} developed at IFAE from which the charge collection and time resolution are calculated.
The charge is computed for each event as the integral of the signal area. For each bias voltage point the charge distribution is fitted with a Landau-Gauss convoluted function and the collected charge is defined as the most probable value of the fit function.
For the time resolution, the Constant Fraction Discriminator (CFD) method is used, taking into account the contribution of the reference sensor, which is $\sigma_{REF}$=\SI{35.7}{\pico\second}, calculated from previous calibration measurements~\cite{Lucia}. The $\sigma_{DUT}$ is obtained and presented as a function of the bias voltage for different neutron fluences.

\subsection{Si-on-Si wafer}
\begin{figure}[h!]
\centering
\subfigure
	{\includegraphics[width=0.45\textwidth]{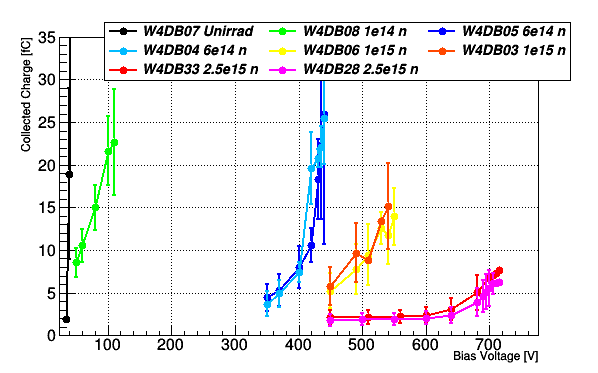}
	} \qquad
\subfigure
	{\includegraphics[width=0.45\textwidth]{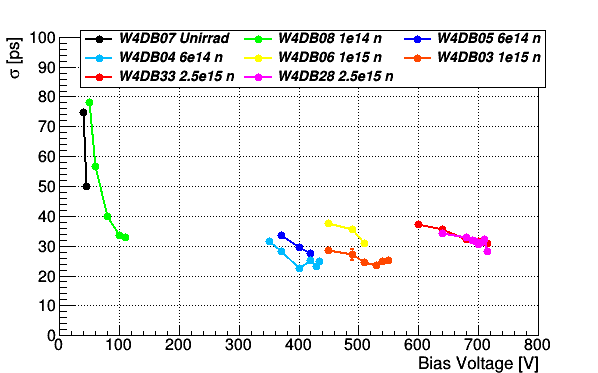}
} \qquad
\caption{Collected charge (left) and time resolution (right) for Si-on-Si wafer sensors before and after irradiation.}
\label{aida_charge_time}
\end{figure}

The collected charge and time resolution are shown in Figure~\ref{aida_charge_time}.
Sensors irradiated to the highest fluence reach the target\footnote{The \SI{4}{\femto\coulomb} low limit is given by the HGTD sensor and electronics requirements.}  of \SI{4}{\femto\coulomb} for bias voltages higher than \SI{680}{\volt}. All the irradiated sensors achieve time resolutions below \SI{40}{\pico\second}. 
Regarding the unirradiated sensor, this one shows a poor performance in time resolution due to the high rate of self-triggers already at \SI{40}{\volt}, see Figure~\ref{fig:autotriggering} (left). Comparing the time resolution value between the unirradiated sensor with the 10$^{14}$n$_{eq}$/cm$^2$ irradiated sensor, for a fixed value of collected charge, the time resolution on the unirradiated device is poorer than the one for the irradiated sensor. This is due to the high rate of self trigger of the unirradiated sensor, which is, for a bias voltage value of \SI{40}{\volt} and a charge collected of $\sim$\SI{19}{\femto\coulomb}, around \SI{7}{\kilo\hertz}. For the irradiated device, at \SI{100}{\volt} bias voltage with a collected charge of $\sim$\SI{21}{\femto\coulomb}, the rate of self triggers is below \SI{1}{\kilo\hertz}, not affecting the timing performance of this device.

\subsection{Low-resistivity epitaxial wafer}
\begin{figure}[h!]
\centering
\subfigure
	{\includegraphics[width=0.45\textwidth]{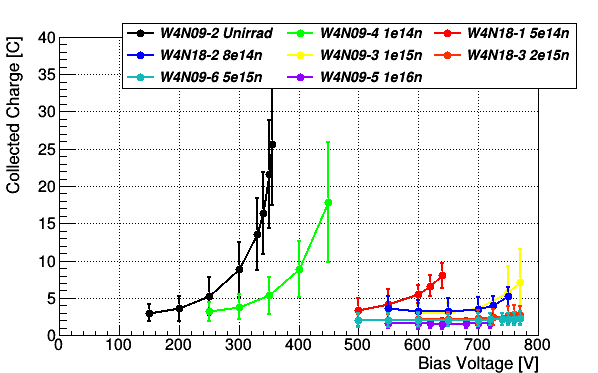}
	} \qquad
\subfigure
	{\includegraphics[width=0.45\textwidth]{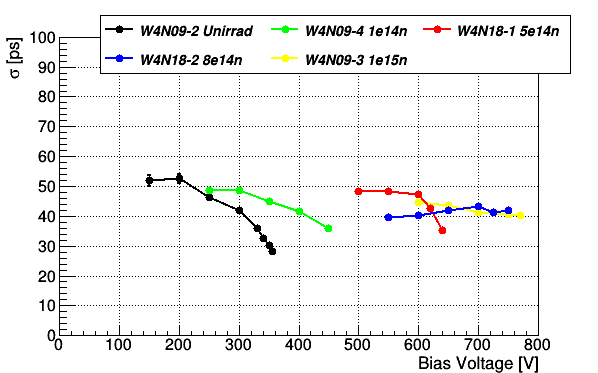}
} \qquad
\caption{Collected charge (left) and time resolution (right) for low-resistivity epitaxial wafer sensors before and after irradiation.}
\label{epi_charge_time}
\end{figure}

The collected charge and time resolution are shown in Figure~\ref{epi_charge_time}. The sensor irradiated at the highest fluence does not show any gain up to \SI{720}{\volt}, but the high leakage current (see Figure~\ref{fig:IVs} (right)) prevented safely reaching higher voltages. The sensor irradiated at 10$^{15}$ n$_{eq}$/cm$^{2}$ reaches a collected charge of \SI{4}{\femto\coulomb} for bias voltages higher than \SI{700}{\volt}. 
Unirradiated and irradiated sensors feature a time resolution around \SI{40}{\pico\second}.

\section{Transient Current Technique}
\subsection{Inter-pad distance for Si-on-Si sensors}
\begin{figure}[h!]
    \centering
\subfigure
    {\includegraphics[width=0.40\textwidth]{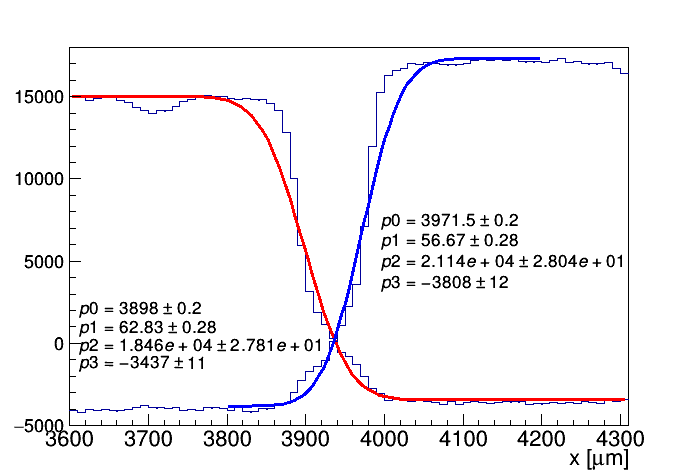}
    }\qquad
\subfigure
    {\includegraphics[width=0.45\textwidth]{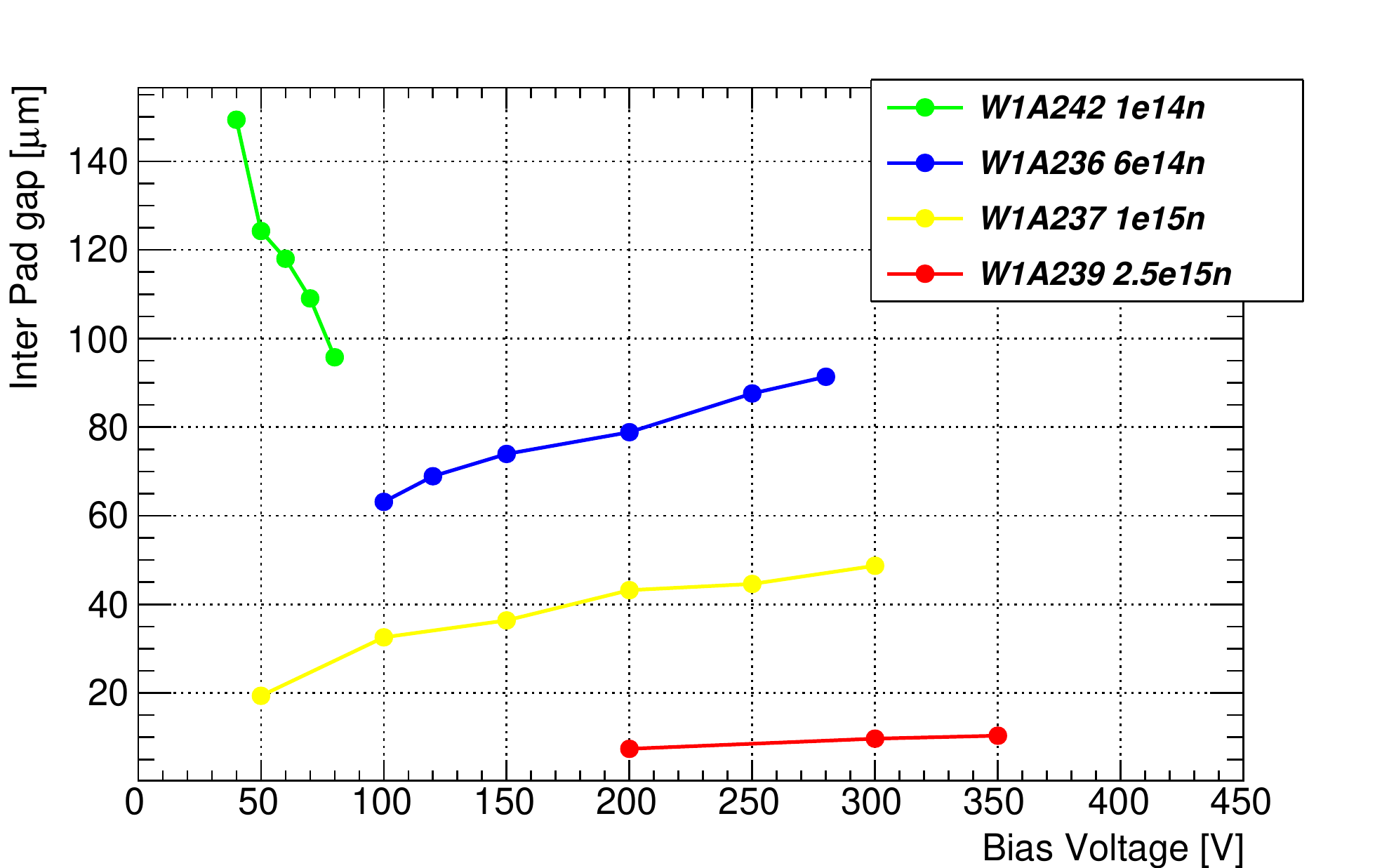}
    }\qquad
\caption{Projections and error function fits of the 2D map for each of the two pads (left) and inter-pad distances (right) measured with TCT for several fluences of Si-on-Si $2\times 2$ LGAD arrays. The difference in collected charge of the two pads in the left figure is related to the non-perpendicular incidence of the infrared light on the entire sensor surface.}
\label{fig:ipgap}
\end{figure}

The inter-pad region of $2\times 2$ LGAD pads of $1.3\times 1.3$ mm$^{2}$ from the Si-on-Si wafer has been studied using the Transient Current Technique (TCT)~\cite{TCT}. 
The nominal gap before irradiation, that is the gap at the sensor design level, is \SI{57}{\micro\metre}.
Four irradiated sensors from 10$^{14}$ to 2.5 $\times$ 10$^{15}$ n$_{eq}$/cm$^{2}$ have been measured at $-20~^{\circ}$C in a dry environment.
An infra-red (IR) laser illuminated the sensor through an opening in the metal layer on the backside of the device. 
A $0.5\times0.5$~mm$^2$ inter-pad area between two pads has been scanned with \SI{50}{\micro\metre} steps in both X-Y directions collecting an average of 1000 waveforms for each laser position. Data have been analyzed offline and a 2D map has been built for each voltage point.
A (reverse) error function is used to fit the 2D map projection distribution for each of the two pad of the DUT, as shown in Figure~\ref{fig:ipgap} (left).
The inter-pad distance is estimated as the difference of the X positions at $50\%$ of the height of the fit function for each of the two pads: IP$_{gap}$=|x$_{50\% PAD_A}$-x$_{50\% PAD_B}$|. 
The IP$_{gap}$ is shown in Figure~\ref{fig:ipgap} (right) as a function of bias voltage.
At low fluences (blue line), part of the carries generated underneath the gain layer drift to the Junction Termination Extension (JTE) and don't produce charge multiplication, the effect is reduced as the bias voltage is increased.
However, at higher fluence some gain is achieved by carries drifting at the JTE-p bulk interface due to field focusing and this results in smaller inter-pad distances than nominal value measured at higher voltages (black line). This behavior was also observed on LGADs from other vendors~\cite{ipgap_study}.


\subsection{Gain and Time resolution for epitaxial sensors}
\begin{figure}[h!]
\centering
\subfigure
	{\includegraphics[width=0.45\textwidth]{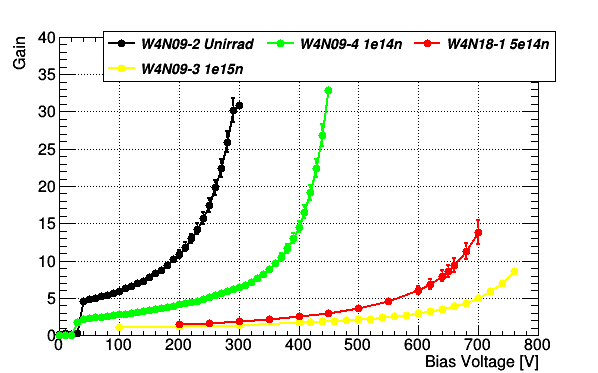}
	} \qquad
\subfigure
	{\includegraphics[width=0.45\textwidth]{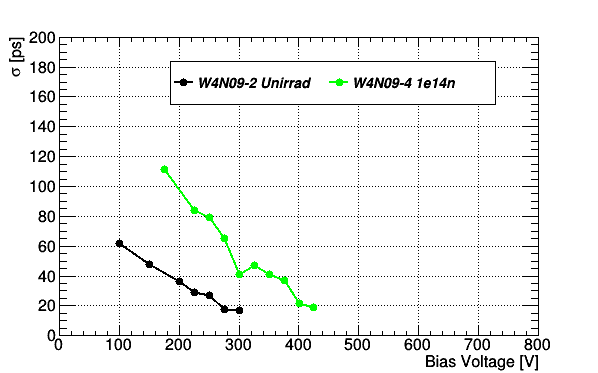}
} \qquad
\caption{Gain (left) and time resolution (right) of the neutron irradiated epitaxial sensors measured with TCT.}
\label{epi_tct}
\end{figure}

Gain measurements have been performed using the TCT at $-20~^{\circ}$C with an IR laser (Figure~\ref{epi_tct} left). Figure~\ref{epi_tct} (right) also shows the time resolution obtained from the time difference of a laser pulse and the same pulse delayed by \SI{50}{\nano\second}. These preliminary results are in agreement within the errors (not reported in the plot) with results measured with a Sr-90 radioactive source shown in Figure~\ref{epi_charge_time}, considering the use of a different CFD fraction (usually $> 50\%$ for TCT). 

\section{Conclusion}
Measurements of Si-on-Si devices show that the target charge of \SI{4}{\femto\coulomb} can be achieved up to 2.5 $\times$ 10$^{15}$ n$_{eq}$/cm$^{2}$ at \SI{700}{\volt}, while epitaxial devices require a higher voltage of \SI{750}{\volt} for fluences up to 10$^{15}$ n$_{eq}$/cm$^{2}$.
%
The better performances of Si-on-Si could be explained by the difference of the wafer resistivity, since the low-resistivity (\SI{100}{\ohm\centi\metre}) of the epitaxial wafers requires a higher bias to be operated.
On the other hand, the Si-on-Si devices show a marginal performance before irradiation due to the restricted operational voltage range. 
TCT inter-pad gap measurements show that after sufficient irradiation the effective gaps are lower than the nominal values.
Preliminary results of time resolution using the TCT are in agreement with the ones obtained with the Sr-90 set-up for epitaxial sensors.





\end{document}